%% file: arxiv_submission_gam_nguyen.tex
\input macros_period.tex

\font \smallfont=cmr10 at 8pt
\font \bigfont=cmr10 at 16pt

\tolerance = 10000
\overfullrule=0pt

\footline{\smallfont \hss \folio\hss} 

\centerline{\bigfont \bigmath The period of $x^h + x + 1$ over GF(2)}
\bigskip
\centerline{Gam~D.~Nguyen}
\medskip
\centerline{\it \smallfont Institute for Globally Distributed Open Research and Education}
\centerline{\smallfont gam.nguyen@igdore.org}
\bigskip

\remark{Abstract.} The periods of polynomials can be used to characterize discrete structures such as algebraic error control codes and feedback shift registers.
We study trinomial $x^h+x+1$ over GF(2), which has the maximum number of consecutive zero coefficients and leads to efficient implementation.
Existing results typically deal with {\it finite} values of $h$ and rely on computer computation methods for finding the periods.
In contrast, here we derive closed-form expressions for the periods of this trinomial for {\it infinite} sets of $h$ values. 

\Section {1}{Introduction} 
Polynomials play important roles in many areas of mathematics and science, such as abstract algebra, number theory, error control coding, shift register sequences, cryptography, and random number generation [\Ber, \BrZ, \Gol, \KlK, \LiN, \LiC, \MaS, \Ngu].
Here we consider binary polynomials that have coefficients~0 and~1. Thus, all polynomial operations are performed in the binary field GF(2),  by using polynomial arithmetic modulo~2. 

An important property of a polynomial is its period [\Ber, \LiN, \LiC, \Ngu], which can be used to characterize discrete structures such as cyclic groups and error control codes.
Let $M(x)$ be a polynomial that has positive degree and $M(0)\not=0$. The {\it period} of $M(x)$ is the smallest positive integer $e$ for which $M(x)$ divides $x^e+1$. The period of a polynomial is also called its {\it order} or its {\it exponent} [\LiN].
 
As an application, a binary cyclic code generated by a polynomial $M(x)$ can correct all single errors if its length is bounded above by the period of $M(x)$ [\Ber, \LiC, \MaS, \Ngu]. 
The periods of polynomials for extreme cases can be easily determined. In particular, the period of binomial polynomial $x^h+1$ is $e = h$, and the period of all-one polynomial
$x^h + x^{h-1} + \cdots + x+1$ is $e=h+1$. 
It can be shown that the period $e$ of a general polynomial of degree $h$ is bounded by $h \le e \le 2^h - 1$. A polynomial of degree $h$ is called primitive if it has the maximum period of $2^h-1$ [\Ber, \Gol, \LiN]. 

Computer computation of the period of a general polynomial involves factorization of the polynomial into irreducible polynomials, and factorization of some related integers into primes, as well as other relevant operations such as testing and checking [\Ber, \BrZ, \LiN].

Trinomial $x^h+x^g+1$, where $0<g<h$, and its applications form a popular topic of study [\Ber, \BrZ, \LiN]. 
In this paper, we focus on the case of $g=1$. That is, we study trinomial $x^h+x+1$, 
which is specially interesting, because it has the maximum number of consecutive zero coefficients and leads to efficient implementation [\Ngu, \Zie]. 
The period of $x^h+x+1$ for some $h \le 1000$ is given in [\Zie].
Furthermore, $x^h+x+1$ can be primitive, e.g., when $h = 2, 3, 4, 6, 7, 15, 22, 60, 63, 127, 153, 471, 532, 865, 900$ (see [\Zie]).

Existing results typically deal with {\it finite} values of $h$ and rely on computer computation methods for finding the periods.  
In contrast, here we allow an {\it infinite} number of $h$ values and provide explicit formulas for the resulting period $e$ of $x^h+x+1$. 
In particular, we show that $e=h^2-h+1$ iff $h\in\{2^m+1 : m\ge 1\}$, and $e=h^2 - 1$ iff $h\in\{2^m : m\ge 1\}$. Furthermore, $e \ge h^2 - h + 1$ if $h$ is odd, and $e \ge h^2 - 1$ if $h$ is even.

\remark{Remark\ri~(Notation).}
If $A(x)$ and $M(x)$ are polynomials, then $\MOD{A(x)}{M(x)}$ denotes the remainder polynomial that is obtained when $A(x)$ is divided by $M(x)$. We have ${\rm degree}(\MOD{A(x)}{M(x)}) < {\rm degree}(M(x))$. 
Thus, the period of $M(x)$ is the smallest positive integer $e$ for which $\MOD{x^e}{M(x)}=1$. 

\remark{Remark\rii.} 
For any polynomials $C(x)$, $D(x)$, $K(x)$, and $M(x)$, the following 2 equalities can be verified
$$
\MOD{C(x)D(x)}{M(x)}=\MOD{\MOD{C(x)}{M(x)}\MOD{D(x)}{M(x)}}{M(x)} 
$$
$$
\MOD{C(x)}{M(x)} = \MOD{C(x)+K(x)M(x)}{M(x)}	
$$
Using the above equalities, for any $i, j \ge 0$, it can be shown that 
$$
\eqalignno
{
\MOD{C(x)^{ij}}{M(x)} &= \MOD{(\MOD{C(x)^i}{M(x)})^j}{M(x)}	\cr
					&= \MOD{(\MOD{C(x)^i+K(x)M(x)}{M(x)})^j}{M(x)} \cr
					&= \MOD{(C(x)^i+K(x)M(x))^j}{M(x)} 
					\cr
}
$$
 
\Section {2}{Main results}
Let $e$ be the period of $T(x)=x^h+x+1$. The following theorems (proved later in this paper) give bounds on $e$ and the exact formulas for $e$ for any $h$ that lies in the 2 infinite sets: $\{2^m + 1 : m\ge 0\}$ and $\{2^m : m\ge 1\}$.

\remark{Theorem\td.} 
$e \ge h^2-h+1$ for all $h\ge 2$.
 
\remark{Theorem\tdd.} 
$e=h^2 -h+1$ iff $h\in\{2^m + 1 : m\ge 0\}$.  

\remark{Theorem\tf.}   
$e \ge h^2-1$ for all even $h \ge 2$.  

\remark{Theorem\tff.}   
$e = h^2-1$ iff $h\in\{2^m : m\ge 1\}$. 
 
\bigskip

Theorems\td~and\tdd~show that 
the period $e$ of $x^h+x+1$ is bounded below by $h^2-h+1$, which is achieved  iff $h = 2^m+1$ for some $m\ge 0$. Note that $h = 2^m+1$ is odd when $m\ge 1$. Theorems\tf~and\tff~show that the lower bound for the period $e$ increases from $h^2-h+1$ to $h^2-1$ when $h$ is constrained to be even. 
By combining the above 4 theorems, we obtain the following theorem.

\remark{Theorem\th.} 
If $h$ is odd, then  $e \ge h^2-h+1$ and $e = h^2-h+1$ iff $h\in\{2^m + 1 : m\ge 1\}$. If $h$ is even, then  $e \ge h^2-1$ and  $e = h^2-1$  iff $h\in\{2^m : m\ge 1\}$.

\remark{Remark\riv.} 
Let $e$ be the period of $x^h+x+1$. 
The upper bound on period is $2^h-1$ (when this trinomial is primitive).
For the lower bounds in Theorems\td~and\tf, we have  $h^2-1 > h^2-h+1$ iff $h > 2$, and $h^2-1 = h^2-h+1$ iff $h=2$. 
We have the following special cases. 
When $h=2$, we have $e=2^h-1=h^2-1=h^2-h+1=3$.
When $h=3$, we have $h^2-1=8>e=2^h-1=h^2-h+1=7$.
When $h=4$, we have $e=2^h-1=h^2-1=15>h^2-h+1=13$. Thus, the upper and lower bounds for the period coincide for small values of $h$.
For larger $h$, we have $2^h-1 \ge e \ge h^2-h+1$ for odd $h \ge 5$, and $2^h-1 \ge e \ge h^2-1$ for even $h \ge 6$. 

\remark{Remark\rvi.}
The reciprocal polynomial of  polynomial $M(x)$ of degree $h$ is the polynomial $M^*(x)=x^hM(x^{-1})$, which has the same period as that of $M(x)$ [\LiN]. Thus, the theorems in this section also hold for  $T^*(x)=x^h+x^{h-1}+1$, which is the reciprocal polynomial of $T(x)=x^h+x+1$.

\Section {3}{Proof of Theorem\td} 
Let $h \ge  2$, and let 
$e$ be the period of $T(x)=x^h+x+1$. Note that $h^2-h+1 = h(h-1)+1$. 
Assume that $1 \le i\le h(h-1)$.
Below we show that $\MOD{x^i}{T(x)} \not= 1$, which implies that $e \ge h(h-1)+1$. 

We can write $i=qh+r$, where $q \ge 0$ and $0\le r < h$.
We then have
$$
\eqalign
{\MOD{x^i}{T(x)} 
&= \MOD{x^{qh}x^r}{T(x)} \cr
&= \MOD{(x^h + T(x))^q x^r}{T(x)} \cr
}
$$
by using Remark\rii. 
Thus, 
$$
\MOD{x^i}{T(x)} = \MOD{(x+1)^q x^r}{T(x)} = \MOD{\sum_{j=0}^q {q \choose j}x^{j+r}}{T(x)}	\eqno(\ec)
$$
where $1\le i = qh+r \le h(h-1)$.

From the assumption $1 \le i\le h(h-1)$, we have $q \le h-1$. Both $0 \le r \le h-1$ and $q \le h-1$ imply that $q+r \le 2h-2$. There are 3 cases to consider: (1) $q=0$, (2) $r=0$, and (3) both $q, r>0$.

\noindent (1) Case: $q=0$. 

We then have $r=i \ge 1$. Because $1 \le r \le h-1$, we have
$\MOD{x^i}{T(x)} = \MOD{x^r}{T(x)} = x^r \not=1$.

\noindent (2) Case: $r=0$. 

Because $1 \le q \le h-1$, it follows  from (\ec) that 

$\MOD{x^i}{T(x)} 
= \MOD{(x+1)^q}{T(x)} 
= (x+1)^q \not=1$. 

\noindent (3) Case: $q>0$ and $r>0$. 
 
There are 3 further cases to consider: $q+r<h$, $q+r=h$, and $q+r > h$.

\noindent (3.1) Case: $q+r<h$.

It follows from (\ec) that $\MOD{x^i}{T(x)} = (x+1)^q x^r \not= 1$ (because $q\ge 1$ and $r\ge 1$).

\noindent (3.2) Case: $q+r=h$.

We have 
$$
\eqalign
{
(x+1)^q x^r 
&= \sum_{j=0}^q {q \choose j}x^{j+r}	\cr
&= x^r + \sum_{j=1}^{q-1} {q \choose j}x^{j+r}	 + x^{q+r} \cr
}
$$
Because $\MOD{x^{q+r}}{T(x)} = \MOD{x^{h}}{T(x)}= x+1$, we have
$$
\MOD{(x+1)^q x^r}{T(x)} = x^r + \sum_{j=1}^{q-1} {q \choose j}x^{j+r}	 + x+1	\eqno(\ed)
$$
with $q+r=h$. 
There are 2 further cases to consider: $r=1$ and $r \ge 2$.

\noindent (3.2.1) Case: $r=1$. 

Then $q=h-1$. Thus, $h(h-1)\ge i=qh+r=(h-1)h+1$, which is a contradiction. Thus, this case is impossible.

\noindent (3.2.2) Case: $r \ge 2$.

It follows from (\ed) that 
$$
\MOD{(x+1)^q x^r}{T(x)} 
= 1+x+x^r+ \sum_{j=1}^{q-1} {q \choose j}x^{j+r} \not= 1	
$$

\noindent (3.3) Case: $q+r>h$.

From (\ec), we have
$$
\eqalign
{
\MOD{x^i}{T(x)} 
&= \MOD{\sum_{j=0}^{h-r-1} {q\choose j}x^{j+r} + \sum_{j=h-r}^q {q\choose j}x^{j+r}}{T(x)}		\cr
&= \sum_{j=0}^{h-r-1} {q\choose j}x^{j+r} + \MOD{\sum_{j=h-r}^q {q\choose j}x^h x^{j+r-h}}{T(x)} \cr
}
$$
Using Remark\rii, we have  
$$
\eqalign
{
\MOD{\sum_{j=h-r}^q {q\choose j}x^h x^{j+r-h}}{T(x)}
&= \MOD{\sum_{j=h-r}^q {q\choose j}(x^h+T(x)) x^{j+r-h}}{T(x)}	\cr
&= \MOD{\sum_{j=h-r}^q {q\choose j}(x+1) x^{j+r-h}}{T(x)}	\cr
&= \sum_{j=h-r}^q {q\choose j}(x+1) x^{j+r-h}
}
$$
because $q+r-h+1 \le r<h$. We then have 
$$
\MOD{x^i}{T(x)} = \sum_{j=0}^{h-r-1} {q\choose j}x^{j+r} + \sum_{j=h-r}^q {q\choose j}(x+1) x^{j+r-h}	\eqno(\ef)
$$ 
Because $q \le h-1$ and $q+r>h$, we have $r \ge 2$.
There are 2 further cases to consider: $q<h-1$ and $q=h-1$.

\noindent (3.3.1) Case: $q<h-1$. 

Using (\ef), we have 
$$
\MOD{x^i}{T(x)} 
= x^r + \sum_{j=1}^{h-r-1} {q\choose j}x^{j+r} + \sum_{j=h-r}^q {q\choose j}(x+1) x^{j+r-h} 
$$
The degree of the second summation is at most $q+r-h+1$, which is less than $r$ (because $q<h-1$). Thus, $\MOD{x^i}{T(x)}$ includes the term $x^r$. We then have $\MOD{x^i}{T(x)}\not= 1$, because $r\ge 2$.

\noindent (3.3.2) Case: $q=h-1$.

We then have $h(h-1) \ge i=qh+r=(h-1)h+r \ge (h-1)h+1$, which is a contradiction. Thus, this case is impossible.

In summary, when $h\ge 2$, we have $\MOD{x^i}{T(x)} \not= 1$ for $1 \le i\le h(h-1)$. Thus, $e \ge h(h-1)+1=h^2-h+1$.~\QED

\Section {4}{Proof of Theorem\tdd}
Assume that $h \ge 2$.
Note that $h^2-h+1=h(h-1)+1$.
In the following we show that the period of $T(x)=x^h+x+1$ is $h(h-1)+1$ iff $h = 2^m+1$ for some integer $m\ge 0$. Note that Theorem\tdd~is valid for the case of $h=2$ (with $m=0$), because the period of $x^2+x+1$ is 3. Thus, below we only consider the case of $h\ge 3$ (with $m \ge 1$).   
Using Remark\rii, we have 
$$
\eqalign
{\MOD{x^{h(h-1)+1}}{T(x)} 
&= \MOD{x^{h(h-1)}x}{T(x)} \cr
&= \MOD{(x^h + T(x))^{h-1}x}{T(x)} \cr
&= \MOD{(x+1)^{h-1}x}{T(x)} \cr
&= \MOD{\sum_{j=0}^{h-1}{h-1 \choose j}x^{j+1}}{T(x)} \cr
}
$$
Because $h\ge 3$, we can write
$$
\eqalign
{\MOD{x^{h(h-1)+1}}{T(x)}
&= \MOD{x+\sum_{j=1}^{h-2}{h-1 \choose j}x^{j+1}+x^h}{T(x)} \cr
&= \MOD{x+x^h}{T(x)}+\MOD{\sum_{j=1}^{h-2}{h-1 \choose j}x^{j+1}}{T(x)} \cr 
&= \MOD{x+x^h}{T(x)}+\sum_{j=1}^{h-2}{h-1 \choose j}x^{j+1} \cr
}
$$
Because $\MOD{x+x^h}{T(x)}=1$, we have
$$
\MOD{x^{h(h-1)+1}}{T(x)} = 1+\sum_{j=1}^{h-2}{h-1 \choose j}x^{j+1} 	\eqno(\eg)
$$

From Theorem\taii, $h-1 \choose j$ is even for all $1\le j\le h-2$ iff $h-1 \in \{2^m : m \ge 1 \}$ iff $h\in \{2^m+1 : m \ge 1 \}$. 
Thus, $h\in \{2^m+1 : m \ge 1 \}$ iff 
$\sum_{j=1}^{h-2}{h-1 \choose j}x^{j+1}=0$.
Using (\eg), we have  
$\sum_{j=1}^{h-2}{h-1 \choose j}x^{j+1}=0$  
iff $\MOD{x^{h(h-1)+1}}{T(x)}=1$. Thus, 
$$
h\in \{2^m+1 : m \ge 1 \}  
~~{\rm iff}~~ \MOD{x^{h(h-1)+1}}{T(x)}=1	\eqno(\eh)
$$

Recall that $e$ is the period of $T(x)$. 
If $e= h(h-1)+1$, then $\MOD{x^{h(h-1)+1}}{T(x)}=1$. Using (\eh), we then have $h\in \{2^m+1 : m \ge 1 \}$.
Conversely, if $h\in \{2^m+1 : m \ge 1 \}$, then it follows from (\eh) that $\MOD{x^{h(h-1)+1}}{T(x)}=1$. Thus, $e\le h(h-1)+1$.  We also have $e\ge h(h-1)+1$ by Theorem\td. 
Thus, $e= h(h-1)+1$. 

In summary, $e= h(h-1)+1$ iff $h\in \{2^m+1 : m \ge 1 \}$. 
The proof is complete by noting that $h(h-1)+1 = h^2 - h + 1$.~\QED

\Section {5}{Proof of Theorem\tf}
Assume that $h$ is even and $h\ge 2$. Let $e$ be the period of $T(x)=x^h+x+1$. Let $1 \le i\le h^2-2$. Below we show that $\MOD{x^i}{T(x)} \not= 1$, which implies that $e \ge h^2-1$.
Note that Theorem\tf~is valid for the case of $h=2$, because the period of $x^2+x+1$ is 3. Thus, below we only consider the case of $h\ge 4$.

We can write $i=qh+r$, where $q \ge 0$ and $0\le r < h$. Using Remark\rii, we have
$\MOD{x^i}{T(x)} 
= \MOD{x^{qh}x^r}{T(x)} 
= \MOD{(x^h + T(x))^q x^r}{T(x)}$.
Thus,
$$
\MOD{x^i}{T(x)} = \MOD{(x+1)^q x^r}{T(x)}	\eqno(\ei)
$$
From the assumption $1 \le i\le h^2-2$, we have $q \le h-1$. Both $0 \le r \le h-1$ and $q \le h-1$ imply that $q+r \le 2h-2$. There are 3 cases to consider: $q=0$, $r=0$, and $q, r>0$.

\noindent (1) Case: $q=0$. 

We then have $r=i \ge 1$. Because $1 \le r \le h-1$, we have
$\MOD{x^i}{T(x)} = \MOD{x^r}{T(x)} = x^r \not=1$.

\noindent (2) Case: $r=0$. 
 
Because $1 \le q \le h-1$, it follows from (\ei) that
$$
\eqalign
{
\MOD{x^i}{T(x)} 
&= \MOD{(x+1)^q}{T(x)} \cr
&= (x+1)^q \not=1 \cr
}
$$

\noindent (3) Case: $q>0$ and $r>0$. 

There are 3 further cases to consider: $q+r<h$, $q+r=h$, and $q+r>h$.

\noindent (3.1) Case: $q+r<h$.

It follows from (\ei) that $\MOD{x^i}{T(x)} = (x+1)^q x^r \not= 1$ (because $q\ge 1$ and $r\ge 1$).

\noindent (3.2) Case: $q+r=h$.

There are 2 further cases to consider: $q=1$ and $q \ge 2$. 

\noindent (3.2.1) Case: $q=1$. 

Then $r=h-1$ and 
$(x+1)^q x^r = (x+1)x^{h-1} = x^h+x^{h-1}$. From (\ei), we have 
$$
\eqalign
{
\MOD{x^i}{T(x)} 
&= \MOD{(x+1)^q x^r}{T(x)}	\cr 
&= \MOD{x^h+x^{h-1}}{T(x)}	\cr
&= \MOD{x^h}{T(x)} + x^{h-1}	\cr
&= 1 + x + x^{h-1}	\not=1	\cr
}
$$
because $h \ge 4$.

\noindent (3.2.2) Case: $q \ge 2$.

Because $q-1 \ge 1$, we can write
$$
(x+1)^q x^r 
= \sum_{j=0}^q {q \choose j}x^{j+r}	
= x^r + \sum_{j=1}^{q-1} {q \choose j}x^{j+r}	 + x^{q+r} 	\eqno(\ej)
$$
Because $\MOD{x^{q+r}}{T(x)} = \MOD{x^{h}}{T(x)}= x+1$, it follows from (\ei) and (\ej) that 
$$
\MOD{x^i}{T(x)}=\MOD{(x+1)^q x^r}{T(x)} = x^r + \sum_{j=1}^{q-1} {q \choose j}x^{j+r}	 + x+1	\eqno(\ek)
$$
where $q \ge 2$.

There are 2 further cases to consider: $r=1$ and $r \ge 2$. 

\noindent (3.2.2.1) Case: $r=1$. 

Recall that $h\ge 4$. Then $q=h-r=h-1\ge 3$. Using $r=1$ in (\ek), we have
$$
\eqalign
{
\MOD{(x+1)^q x^r}{T(x)} 
&= x^r + \sum_{j=1}^{q-1} {q \choose j}x^{j+r}	 + x+1 \cr
&= x + \sum_{j=1}^{q-1} {q \choose j}x^{j+1}	 + x+1 \cr
&= 1 + \sum_{j=1}^{q-1} {q \choose j}x^{j+1}  \cr
}
$$
Recall that $h$ is even and $h\ge 4$. Then $q=h-1$ is odd and $q\ge 3$. Thus, $q\not=2^m$ for any integer $m\ge 1$. 

Because $q\not=2^m$, Theorem\taii \ implies that there exists $1\le i\le q-1$ such that ${q \choose i}$ is odd. Thus,
$$
\eqalign
{
\MOD{x^i}{T(x)}
&=\MOD{(x+1)^q x^r}{T(x)} 	\cr
&= 1 + \sum_{j=1}^{q-1} {q \choose j}x^{j+1}  \cr
&= 1 + x^{i+1}+\sum_{j=1, j\not=i}^{q-1} {q \choose j}x^{j+1} \not=1 \cr
}
$$

\noindent (3.2.2.2) Case: $r \ge 2$.

Recall that $q+r=h$ for this case. From (\ek), we have  
$$
\eqalign
{
\MOD{(x+1)^q x^r}{T(x)} 
&= x^r + \sum_{j=1}^{q-1} {q \choose j}x^{j+r}	 + x+1	\cr
&= 1+x+x^r+ \sum_{j=1}^{q-1} {q \choose j}x^{j+r} \not=1		\cr
}
$$
because $r\ge 2$.

\noindent (3.3) Case: $q+r>h$.

Then $q>h-r$. Recall that $h \ge r+1$, i.e., $h-r-1 \ge 0$. From (\ei), we have
$$
\eqalign
{
\MOD{x^i}{T(x)} 
&= \MOD{(x+1)^q x^r}{T(x)}		\cr
&= \MOD{\sum_{j=0}^q {q\choose j}x^{j+r}}{T(x)}		\cr
&= \MOD{\sum_{j=0}^{h-r-1} {q\choose j}x^{j+r} + \sum_{j=h-r}^q {q\choose j}x^{j+r}}{T(x)}		\cr
&= \sum_{j=0}^{h-r-1} {q\choose j}x^{j+r} + \MOD{\sum_{j=h-r}^q {q\choose j}x^h x^{j+r-h}}{T(x)} \cr
}
$$
Using Remark\rii, we have 
$$
\eqalign
{
\MOD{\sum_{j=h-r}^q {q\choose j}x^h x^{j+r-h}}{T(x)} 
&= \MOD{\sum_{j=h-r}^q {q\choose j}(x^h+T(x)) x^{j+r-h}}{T(x)}	\cr
&= \MOD{\sum_{j=h-r}^q {q\choose j}(x+1) x^{j+r-h}}{T(x)}	\cr
}
$$
Recall that $q+r \le 2h-2$, which implies that $1+q+r-h < h$. We then have
$\MOD{\sum_{j=h-r}^q {q\choose j}(x+1) x^{j+r-h}}{T(x)} = \sum_{j=h-r}^q {q\choose j}(x+1) x^{j+r-h}$. Thus,
$$
\MOD{x^i}{T(x)} 
= \sum_{j=0}^{h-r-1} {q\choose j}x^{j+r} + \sum_{j=h-r}^q {q\choose j}(x+1) x^{j+r-h} 	\eqno(\en)
$$
Because $q \le h-1$ and $q+r\ge h+1$, we have $r\ge 2$. There are 2 further cases to consider: $q<h-1$ and $q=h-1$.

\noindent (3.3.1) Case: $q<h-1$.

Recall that $r\le h-1$.

If  $r< h-1$, i.e., $h-r-1\ge 1$, then from (\en) we have 
$$
\eqalign
{
\MOD{x^i}{T(x)} 
&= \sum_{j=0}^{h-r-1} {q\choose j}x^{j+r} + \sum_{j=h-r}^q {q\choose j}(x+1) x^{j+r-h} \cr
&= x^r + \sum_{j=1}^{h-r-1} {q\choose j}x^{j+r} + \sum_{j=h-r}^q {q\choose j}(x+1) x^{j+r-h} \cr
}
$$

If  $r= h-1$, i.e., $h-r-1=0$, then from (\en) we have
$$
\eqalign
{
\MOD{x^i}{T(x)} 
&= \sum_{j=0}^{h-r-1} {q\choose j}x^{j+r} + \sum_{j=h-r}^q {q\choose j}(x+1) x^{j+r-h} \cr
&= x^r + \sum_{j=h-r}^q {q\choose j}(x+1) x^{j+r-h} \cr
}
$$
The degree of $ \sum_{j=h-r}^q {q\choose j}(x+1) x^{j+r-h}$ is bounded above by $q+r-h+1<r$ (because $q<h-1$). Thus, $\MOD{x^i}{T(x)}$ includes the term $x^r$. Then $\MOD{x^i}{T(x)}\not= 1$, because $r\ge 2$.

\noindent (3.3.2) Case: $q=h-1$.

Because $1\le i = qh+r \le h^2-2$, we have $r<h-1$. Thus, $h-r-1\ge 1$,  for this case.

Recall that $h\ge 4$ and even. Then $q=h-1$ is odd and $q\ge 3$.  Because ${q\choose 1}=q$ is odd, then from (\en) we have 
$$
\MOD{x^i}{T(x)} = x^{r+1}+\sum_{j=0, j\not=1}^{h-r-1} {q\choose j}x^{j+r} + \sum_{j=h-r}^q {q\choose j}(x+1) x^{j+r-h}
$$ 
The degree of $ \sum_{j=h-r}^q {q\choose j}(x+1) x^{j+r-h}$ is bounded above by $q+r-h+1=r$ (because $q=h-1$). Thus, $\MOD{x^i}{T(x)}$ includes the term $x^{r+1}$. Then $\MOD{x^i}{T(x)}\not= 1$, because $r\ge 2$.

In summary, if $h\ge 2$ is even, we have $\MOD{x^i}{T(x)} \not= 1$ for $1 \le i\le h^2-2$, which implies that $e \ge h^2-1$.~\QED

\Section {6}{Proof of Theorem\tff}
Assume that $h$ is even and $h\ge 2$.
Theorem\tff~is valid for the case of $h=2$ (with $m=1$), because the period of $x^2+x+1$ is 3. Thus, below we only consider the case of $h\ge 4$. 
Let $T(x)=x^h+x+1$. Because $h^2 - 1 = h(h-1) + (h-1)$, we have $x^{h^2-1} = x^{h(h-1)}x^{h-1}$. 
Using Remark\rii, we then have   
$$
\eqalign
{\MOD{x^{h^2-1}}{T(x)} 
&= \MOD{x^{h(h-1)}x^{h-1}}{T(x)} \cr
&= \MOD{(x^h + T(x))^{h-1}x^{h-1}}{T(x)} \cr
&= \MOD{(x+1)^{h-1}x^{h-1}}{T(x)} \cr
}
$$
Thus, 
$$
\eqalign
{\MOD{x^{h^2-1}}{T(x)} 
&= \MOD{\sum_{j=0}^{h-1}{h-1 \choose j}x^{j+h-1}}{T(x)} \cr
&= x^{h-1}+\MOD{\sum_{j=1}^{h-1}{h-1 \choose j}x^{j+h-1}}{T(x)} \cr
&= x^{h-1}+\MOD{x^h\sum_{j=1}^{h-1}{h-1 \choose j}x^{j-1}}{T(x)} \cr
&= x^{h-1}+\MOD{(x^h+T(x))\sum_{j=1}^{h-1}{h-1 \choose j}x^{j-1}}{T(x)} \cr
}
$$
by using Remark\rii. 
We then have
$$
\eqalign 
{\MOD{x^{h^2-1}}{T(x)} 
&= x^{h-1}+\MOD{(x+1)\sum_{j=1}^{h-1}{h-1 \choose j}x^{j-1}}{T(x)} \cr
&= x^{h-1}+(x+1)\sum_{j=1}^{h-1}{h-1 \choose j}x^{j-1} \cr
&= x^{h-1}+\sum_{j=1}^{h-1}{h-1 \choose j}(x^j + x^{j-1}) \cr
}
$$
Because $h \ge 4$, we can write
$\MOD{x^{h^2-1}}{T(x)}
= x^{h-1}+(x^{h-1} + x^{h-2})+\sum_{j=1}^{h-2}{h-1 \choose j}(x^j + x^{j-1})$.
Thus,
$$
\MOD{x^{h^2-1}}{T(x)}
= x^{h-2}+\sum_{j=1}^{h-2}{h-1 \choose j}(x^j + x^{j-1})	\eqno(\ep)
$$
From (\ep), we have 
$\MOD{x^{h^2-1}}{T(x)} 
= x^{h-2}+\sum_{j=1}^{h-2}{h-1 \choose j}x^j + \sum_{j=1}^{h-2}{h-1 \choose j} x^{j-1}$. Because $h \ge 4$, we can write
$$
\eqalign 
{\MOD{x^{h^2-1}}{T(x)}
&= x^{h-2}+\sum_{j=1}^{h-2}{h-1 \choose j}x^j + {h-1 \choose 1}x^0+\sum_{j=2}^{h-2}{h-1 \choose j} x^{j-1}	\cr
&= x^{h-2}+\sum_{j=1}^{h-3}{h-1 \choose j}x^j +{h-1 \choose {h-2}}x^{h-2}+ {h-1 \choose 1}x^0+\sum_{j=1}^{h-3}{h-1 \choose j+1} x^{j}	\cr
}
$$
Thus,
$$
\MOD{x^{h^2-1}}{T(x)}
= {h-1 \choose 1}x^0+\sum_{j=1}^{h-3}\left[{h-1 \choose j}+{h-1 \choose j+1}\right] x^{j}	+{h-1 \choose {h-2}}x^{h-2}	+x^{h-2}	\eqno(\eq)
$$
From (\eq), we have 
$\MOD{x^{h^2-1}}{T(x)} = 1$ iff   
${h-1 \choose j}$ is odd for all $1 \le j \le h-2$. Furthermore, Remark\rn~states that ${h-1 \choose j}$ is odd for all $1\le j \le h-2$ iff $h \in \{2^m : m \ge 1 \}$. Thus,
$$
h \in \{2^m : m \ge 1 \} ~~~{\rm iff}~~~ \MOD{x^{h^2-1}}{T(x)} = 1	\eqno(\ew)
$$

Recall that $e$ is the period of $T(x)=x^h+x+1$.   
If $h \in \{2^m : m \ge 1 \}$, then (\ew) implies that $\MOD{x^{h^2-1}}{T(x)}=1$. Thus, $e\le h^2-1$. We also have $e\ge h^2-1$ by Theorem\tf. 
Thus, $e=h^2-1$. 
Conversely, if $e=h^2-1$, then $\MOD{x^{h^2-1}}{T(x)}=1$. Then (\ew) implies that $h \in \{2^m : m \ge 1 \}$.
In summary, $e=h^2-1$ iff $h \in \{2^m : m \ge 1 \}$.~\QED

\bigskip

\Section{7}{Appendixes}
The results presented below are useful for deriving the polynomial periods.

\remark{Remark\riii.} Let $p$ be a prime and $h>0$ be an integer.  Then there are integers $q>0$ and $j\ge 0$ such that $h=qp^j$ and $p$ does not divide $q$. This fact can be proved as follows.  Using the prime-factorization theorem, we have $h=p^j\prod_{i=1}^kp^{n_i}_i$, for some integers $j\ge 0$, $k\ge 1$, $n_i\ge0$, $1\le i\le k$, and distinct primes $p_1\not=p$, $p_2\not=p$, \dots, $p_k\not=p$. Let $q=\prod_{i=1}^kp^{n_i}_i$. Then $q>0$ and $h=qp^j$. Note that $p$ does not divide $q$, because $p_i$ is prime and $p_i\not=p$, $1\le i\le k$.

\remark{Theorem\taii.} Let $p$ be a prime  and let $h\ge 2$ be an integer. Then $p$ divides $h \choose i$ for all $1\le i\le h-1$ iff $h \in \{p^m : m\ge 1 \}$. 
For the special case of $p=2$, we have $ h  \choose i$ is even for all $1\le i\le h-1$ iff $h \in \{2^m : m\ge 1 \}$.

\remark{Proof.} 

\noindent(1) Only-if part: Assume that $p$ divides $h \choose i$ for all $1\le i\le h-1$. Suppose that $h\not=p^m$ for  any integer $m\ge 1$. Using Remark\riii, there are integers $q>0$ and $j\ge 0$ such that $h=qp^j$ and $p$ does not divide $q$. Because $h\not=p^m$ for  any integer $m\ge 1$, we have $q>1$.
Let $I=p^j$. Then $h/I = q$ and
$$
{h\choose I}=\prod^{I-1}_{k=0}\frac{h-k}{I-k}=\frac{h}{I}\prod^{I-1}_{k=1}\frac{h-k}{I-k}=q\prod^{I-1}_{k=1}\frac{h-k}{I-k}
$$

For each $k\in\{1,2,\dots,I-1\}$, by Remark\riii, we have $k=p^{m(k)}s(k)$ for some integers $m(k)\ge 0$ and $s(k)>0$, such that $p$ does not divide $s(k)$. Because $k\le I-1$, we have $p^{m(k)}s(k) \le p^j -1$, which implies that $j>m(k) \ge 0$ for all $k\in\{1,2,\dots,I-1\}$. We then have
$$
\frac{h-k}{I-k} = \frac{qp^j-p^{m(k)}s(k)}{p^j-p^{m(k)}s(k)} =  
\frac{qp^{j-m(k)}-s(k)}{p^{j-m(k)}-s(k)}
$$

We have $1\le I\le h-1$, because $q>1$. From the assumption, $p$ divides ${h\choose I}=q\prod^{I-1}_{k=1}\frac{h-k}{I-k}$. Because $p$ does not divide $q$ and $p$ is prime, $p$ divides 
$$
\prod^{I-1}_{k=1}\frac{h-k}{I-k}=\prod^{I-1}_{k=1}\frac{qp^{j-m(k)}-s(k)}{p^{j-m(k)}-s(k)}$$
which implies that $p$ divides 
$$
\prod^{I-1}_{k=1}\left[qp^{j-m(k)}-s(k)\right]
$$
Thus, $p$ divides $qp^{j-m(K)}-s(K)$ for some $1\le K\le I-1$. Then $pr=qp^{j-m(K)}-s(K)$ for some integer $r>0$. Note that $s(K)>0$ and $j>m(K)$, i.e., $j-m(K)-1\ge 0$. We then have 
$$
s(K)=qp^{j-m(K)}-pr = p\left[qp^{j-m(K)-1}-r\right]
$$
which implies that $p$ divides $s(K)$, which is a contradiction.  Thus, $h=p^m$ for some integer $m\ge 1$.

\noindent (2) If part:  Assume that $h=p^m$ for some integer $m\ge 1$.
Let $1\le i \le h-1=p^m-1$.  From Remark\riii, there are integers $k\ge 0$ and $M>0$ such that $i=p^kM$ and $p$ does not divide $M$. We have $k<m$, because $i \le p^m-1$ and $M>0$.
Note that
$$
{p^m \choose i} = \frac{p^m}{i}{p^m-1 \choose i-1} 
$$

By letting $N={p^m-1 \choose i-1}$, we then have

$$
\eqalign
{
{p^m \choose i} 
&= \frac{p^m}{p^k} \frac{N}{M}	\cr
&= p^{m-k} \frac{N}{M}	\cr
&= p^{j} \frac{N}{M}	\cr
}
$$
where $j=m-k>0$.
Because $\gcd (M, p^{j})=1$, $M$  divides $N$, which implies that ${p^m \choose i}$ is a multiple of $p^{j}$. Thus, $p$ divides ${p^m \choose i}={h \choose i}$.~\QED 

\remark{Theorem\taiv.} Let $p$ be a prime, $m\ge 1$, and $h=p^m$. If $0 \le i \le h-1$, 
then $p$ does not divide $h-1 \choose i$.

\remark{Proof.} 
Let $i \in \{0, 1, \dots, h-1 \}$. From Exercise~1.1 of~[\Aya], we have
${{h-1}\choose i} =  (-1)^i ({\rm mod} \ p)$. Thus, ${{h-1}\choose i} -  (-1)^i = kp$ for some integer $k$.
Suppose that $p$ divides ${{h-1}\choose i}$.
Then ${{h-1}\choose i} = np$ for some integer $n \ge 1$. Thus, $np-  (-1)^i = kp$.
We then have $(n-k)p=(-1)^i$, which is impossible. 
Thus, $p$ does not divide ${{h-1}\choose i}$.~\QED

The converse of Theorem\taiv \ does not hold for arbitrary prime $p$. That is, if $p$ does not divide $h-1 \choose i$ for any $i \in \{0, 1, \dots, h-1 \}$, then it is not necessary that $h=p^m$ for some integer $m\ge 1$.  For example,  let $p=3$ and $h=6$. Then $p$ does not divide $h-1 \choose i$ for any $i \in \{0, 1, \dots, h-1 \}$, but $h\not=p^m$ for any integer $m \ge 1$.
Theorem\tavii~below shows  that the converse of Theorem\taiv~also holds for the special case of $p=2$.

\remark{Theorem\tavii.} Let $h\ge 2$ be an integer. Then 
$ h - 1 \choose i$ is odd for all $0 \le i \le h-1$ iff $h\in\{2^m : m\ge 1\}$.

\remark{Proof.} 

\noindent (1) Assume that $h\in\{2^m : m\ge 1\}$. Using Theorem\taiv~with $p=2$, it follows that 2 does not divide $h-1 \choose i$, i.e., $h-1 \choose i$ is odd, for all $0\le i\le h-1$.

\noindent (2) Let $h\ge 2$ be an integer.  Assume that  $h-1 \choose i$ is odd, for all $0\le i\le h-1$.

In general, for $1\le i\le h-1$, we have
$$
{h-1 \choose i} = {h-1 \choose i-1} \frac{h-i}{i}
$$
From Remark\riii, there are integers $q>0$ and $m\ge 0$ such that $h=q2^m$ and $2$ does not divide $q$, i.e., $q$ is odd.

Suppose that $q\not=1$. Thus, $q\ge 3$.  Let $I=2^m$. We then have 
$$
\frac{h-I}{I} = \frac{q2^m-2^m}{2^m} = q-1
$$
which is even and positive (because $q$ is odd and $q\ge 3$). Note that $1\le I \le h-1$, because $m\ge 0$ and $q\ge 3$. Thus,
$$
{h-1 \choose I} = {h-1 \choose I-1} \frac{h-I}{I}={h-1 \choose I-1}(q-1)
$$
is  even, which contradicts the assumption. Thus, $q=1$, and then $h=2^m$. We have $m\ge 1$, because $h\ge 2$. Thus, $h\in\{2^m : m\ge 1\}$.~\QED
\remark{Remark\rn.} 
Let $h\ge 3$ be an integer. We then have ${{h - 1} \choose {0}} = {{h - 1} \choose {h-1}} = 1$, which is odd.
From Theorem\tavii, we can conclude that  
$h - 1 \choose i$ is odd for all $1 \le i \le h-2$ iff $h\in\{2^m : m\ge 1\}$.

\bigskip

\baselineskip=14 pt

\Section {References}

\noindent{[\Aya]} M.~Ayad, {\it Galois Theory and Applications: Solved Exercises and Problems}, World Scientific, 2018.
 
\noindent{[\Ber]} E.R.~Berlekamp, {\it Algebraic Coding Theory}, McGraw-Hill, 1968.

\noindent{[\BrZ]} R.P.~Brent and P.~Zimmermann, The Great Trinomial Hunt, {\it Notices of the AMS}, vol.~58, no.~2, pp.~233-239, Feb.~2011.

\noindent{[\Gol]} S.W.~Golomb, {\it Shift Register Sequences}, 2nd edition, Aegean Park Press, 1982. 

\noindent{[\KlK]} T.~Klove and V.~Korzhik, {\it Error Detecting Codes: General Theory and their Application in Feedback Communication Systems}, Kluwer Academic, 1995. 

\noindent{[\LiN]} R.~Lidl and H.~Niederreiter, {\it Introduction to Finite Fields and Applications}, 
Cambridge University Press, 1986.

\noindent{[\LiC]} S.~Lin and D.J.~Costello, Jr., {\it Error Control Coding: Fundamentals and Applications}, Prentice Hall, 1983.

\noindent{[\MaS]} F.J.\ MacWilliams and N.J.A.~Sloan, {\it The Theory of Error-Correcting Codes}, North-Holland, 1977.

\noindent {[\Ngu]} G.D.~Nguyen, ``Fast CRCs," {\it IEEE Transactions on Computers}, vol.~58, no.~10, pp.~1321-1331, Oct.~2009.

\noindent{[\Zie]} N.~Zierler, On $x^n + x + 1$ over GF(2), {\it Information and Control}, vol.~16, no.~5, pp.~502-505, 1970.

\vfill\eject\break

\bye

%% file: macros_period.tex
\def \Section #1#2 
{
{
\goodbreak
\bigskip
\par 
\parindent = 0ex 
\bf 
\setbox0=\hbox{#1} \dimen0=\wd0 
\setbox1=\hbox{\ \ \ } \dimen1=\wd1 
\advance \dimen0 by \dimen1 
\hangindent=\dimen0 
\rightskip=0pt plus 1fil
\leftskip=0pt 
{\hyphenpenalty=99999 {#1}\ \ \ {\ignorespaces #2}} 
\bigskip  
\par
}
\noindent 
}

\def \remark #1
{
\bigskip
\noindent \ignorespaces
{{\bf#1}\ }
\nobreak \ignorespaces
}
 
\def\QED{\nobreak \hfill{$\sqcup \kern -0.66em  \sqcap$} \medskip}
\def \frac#1#2{{#1 \over #2}}


\font\variable		=cmmi10	at 16pt	 
\font\scriptvariable=cmmi10 at 11pt 
\font\roman			=cmr10 	at 16pt	

\def\bigmath
{
\textfont0		=\roman				
\textfont1		=\variable 			
\scriptfont1	=\scriptvariable	
}

\def \Aya{1}
\def \Ber{2}	
\def \BrZ{3}
\def \Gol{4}
\def \KlK{5}
\def \LiN{6}
\def \LiC{7}
\def \MaS{8}
\def \Ngu{9}
\def \Zie{10}

\def \td		{~1}
\def \tdd		{~2}
\def \tf		{~3}
\def \tff		{~4}
\def \th		{~5}
\def \taii		{~A1}
\def \taiv		{~A2}
\def \tavii		{~A3}

\def \ec		{1}
\def \ed		{2}
\def \ef		{3}
\def \eg		{4}
\def \eh		{5}
\def \ei		{6}
\def \ej		{7}
\def \ek		{8}
\def \en		{9}
\def \ep		{10}
\def \eq		{11}
\def \ew		{12}

\def \ri      	{~1}
\def \rii      	{~2}
\def \riv     	{~3}
\def \rvi     	{~4}
\def \riii      {~A1} 
\def \rn	    {~A2}

\def \MOD #1#2{{\rm R}_{#2}\left[#1\right]} 